\documentclass{article}
\usepackage{multirow}
\usepackage{amsfonts}
\usepackage{bbm}
\usepackage{stmaryrd}
\usepackage{xcolor}
\usepackage{cite}
\usepackage{comment}
\usepackage[symbol]{footmisc}
\usepackage{spconf,amsmath,graphicx}
\usepackage{url}
\usepackage{subcaption}
\usepackage{multirow}
\usepackage{amsfonts}
\usepackage{bbm}
\usepackage{stmaryrd}
\usepackage{xcolor}
\usepackage{cite}
\usepackage{comment}
\usepackage[symbol]{footmisc}

\title{LC4SV: A Denoising Framework Learning to Compensate for Unseen Speaker Verification Models}
\name{Chi-Chang Lee$^{1,2}$, Hong-Wei Chen$^{1}$, Chu-Song Chen$^{1,2}$, Hsin-Min Wang$^{2}$, Tsung-Te Liu$^{1}$, Yu Tsao$^{2}$}
\address{
$^{1}$National Taiwan University, Taipei, Taiwan\\
$^{2}$Academia Sinica, Taipei, Taiwan}
\copyrightnotice{979-8-3503-0689-7/23/\$31.00~\copyright2023 IEEE}
\begin{document}
%
\maketitle
\begin{abstract}
The performance of speaker verification (SV) models may drop dramatically in noisy environments.
A speech enhancement (SE) module can be used as a front-end strategy.
However, existing SE methods may fail to bring performance improvements to downstream SV systems due to artifacts in the predicted signals of SE models.
To compensate for artifacts, we propose a generic denoising framework named LC4SV, which can serve as a pre-processor for various unknown downstream SV models.
In LC4SV, we employ a learning-based interpolation agent to automatically generate the appropriate coefficients between the enhanced signal and its noisy input to improve SV performance in noisy environments.
Our experimental results demonstrate that LC4SV consistently improves the performance of various unseen SV systems.
To the best of our knowledge, this work is the first attempt to develop a learning-based interpolation scheme aiming at improving SV performance in noisy environments.
\end{abstract}
%

\section{Introduction}
\label{sec:intro}
A speech enhancement (SE) process involves extracting speech components from a distorted speech signal to create an enhanced signal that exhibits better properties~\cite{SE}.
Recently, SE was formulated as a regression task that converts noisy speech signals into clean ones using deep neural network-based mapping functions~\cite{CHLee, SE1, SE2, SE3, Liu2014ExperimentsOD, DAELu, 7422753, hu2020dccrn, seril, nastar}.
Most of the time, signal-level distance measures are used as objective functions for model training (e.g., L1 norm~\cite{pandey2018new}, L2 norm~\cite{1164453}, SI-SDR~\cite{SDR}, or multiple-resolution loss~\cite{demucs_stft}).
For the evaluation of SE performance, a standard measure is to conduct a subjective listening test, usually requiring measurement of speech quality and intelligibility.
However, achieving unbiased subjective assessment results requires a large-scale listening test (a large number of listeners, each listening to a large number of speech utterances), which is often prohibitive.
To overcome this limitation, numerous objective evaluation metrics have been developed~\cite{PESQ, STOI, reddy2021dnsmos, cooper2022generalization}.

On the other hand, the SE unit, as a separately trained pre-processor, can play a key role in speech-related applications, improving the performance without modifying the main application system parameters~\cite{d4am, espnet, 9747146}.
In particular, speaker identification (SID) and speaker verification (SV) tasks are popular speech-related applications.
The model architecture and objective function of SE have been investigated in several studies to improve the performance of SID and SV under noisy conditions~\cite{deto,9064910}.
Moreover, some studies explored that SE-induced artifacts and distortions may even degrade SV performance~\cite{deto}.
To deal with this issue, Shon et al.~\cite{Shon2019VoiceIDLS}, Mošner et al.~\cite{9747771}, and Dowerah et al.~\cite{10022350} directly cascade an SV model and use the corresponding SV target errors instead of signal-level distances (e.g., L1 and L2) to train SE models.
However, most commercial SV systems are often provided by third parties, which may not be accessible when training SE models.
Since different SV models differ in various training settings, an SE unit trained only on the SV objective of one specific SV model may not generalize well to other unseen SV systems.
Directly feeding the predicted output of an SE model into a downstream model would introduce severe artifacts and destroy the integrity of critical information required for SID and SV tasks.
To prevent artifacts and over-parameterization, we employ a learning-based interpolation agent. 
The agent automatically determines an appropriate linear combination of the enhanced output and the corresponding noisy input to compensate for the information, thereby mitigating the adverse effects of the SE model and improving SV performance.

In this work, we design a generic denoising framework, LC4SV, as a ``universal'' pre-processor for SV in noisy environments by combining an SE unit for primary denoising and an interpolation agent as a compensation processor.
First, we pre-train the SE model with a signal-level measurement and then fine-tune it with the SV objective generated by a proxy SV model.
Next, to prevent the enhanced results from lacking crucial information required by another proxy SV model, we additionally train a reinforcement learning (RL)-based interpolation agent to automatically determine the appropriate interpolation of the enhanced output and the noisy input for compensation.
The RL-based interpolation agent is trained to achieve a lower error rate of verification results, so the interpolated output can better generalize to other SV models accordingly.
According to the experimental results, LC4SV is able to improve the SV performance of various unseen SV systems, outperforming the SV performance of the same SV systems using noisy or enhanced speech signals of other SE models.
Furthermore, LC4SV achieves better performance compared to two baseline methods: (i) using the signal-to-noise ratio (SNR) as a hard threshold to select either enhanced or noisy signals to feed into the SV model and (ii) predetermining the interpolation coefficients.
This study makes two main contributions:
(1) to the best of our knowledge, this is the first attempt to develop a generic SE preprocessor applicable to a variety of unseen SV systems;
(2) we propose a rational interpolation scheme for compensation and relate it to the motivation for better generalization.

\section{BACKGROUND and MOTIVATION}
\label{sec:background}
\subsection{Noise Robustness Strategies for Speaker Verification}
It is well known that acoustic mismatch (often caused by noise interference) limits the applicability of speech-related techniques.
To improve noise robustness, most SV models tend to use real or synthetic noisy speech utterances as data augmentation for model training~\cite{chung2020in}.
For example, the well-known freely available speaker recognition dataset Voxceleb~\cite{NAGRANI2020101027} provides a large amount of real noisy speech data. Many SV models benefit from its sufficient training data to achieve good robustness~\cite{zeng2022joint}.
However, most of the SV systems provided by third parties are black-box models whose parameters cannot be fine-tuned to improve performance in specific noisy environments, especially at low SNR.
In contrast, an SE unit can be trained separately and used as an advanced front-end processor combined with existing SV models to form a two-stage process for better noise robustness.
To further improve the effect of SE units on SV, several attempts have been made to directly update SE model parameters by using the prediction error of a specific SV model~\cite{Shon2019VoiceIDLS, 9747771}.
In their scenarios, the SV model used for training and evaluation is the same. 
However, when the evaluated SV system is different from the trained system, the SV performance with enhanced speech utterances would be sub-optimal due to the remained artifacts and the over-parameterization issue.

\subsection{Artifacts Caused by SE Models}
In general, SV models exhibit robustness to natural noisy speech signals because they are usually included in the training data as part of the data augmentation scheme.
In contrast, artifacts in the output of certain SE models may have unseen effects on SV models. 
Given the limited model capacity, SE models inevitably introduce artifacts, which can degrade the performance of downstream SV models for reasons that are difficult to trace.

Constrained by the inability to simultaneously modify the front-end and back-end models, it is necessary to devise a post-processing strategy to address the artifacts.
An effective and straightforward solution is to employ interpolation techniques instead of directly feeding the enhanced speech signals into the SV model.
By combining the enhanced signal with its corresponding noisy input, we impose a constraint to ensure that the resulting signal remains close to the characteristics of natural signal.
This approach helps to compensate for lost information and mitigate the negative effects of artifacts.
Recently, in the field of Automatic Speech Recognition (ASR), Sato et al.~\cite{enh_not} proposed a learning-based method to predict the likelihood of an enhanced signal degrading the performance compared to the original noisy signal.
Instead of treating the possibility as a hard threshold, they turned it into an interpolated weight value and empirically improved ASR performance by using this interpolation scheme.
However, unlike the ASR task, which calculates the performance metric from individual samples, the performance metric of SV needs to be determined from multiple paired samples.
Our framework is proposed to construct an interpolation scheme directly aimed at maximizing SV performance.
By pairwisely considering the contrasting performance within a batch, we develop an RL-based method to optimize the SV metric by taking appropriate interpolation coefficients as actions.
In Sec.~\ref{sec:method}, we further illustrate the details of our proposed framework.

\label{sec:method}
\begin{figure}[h]
        \centering
        \includegraphics[width=0.33\textwidth]{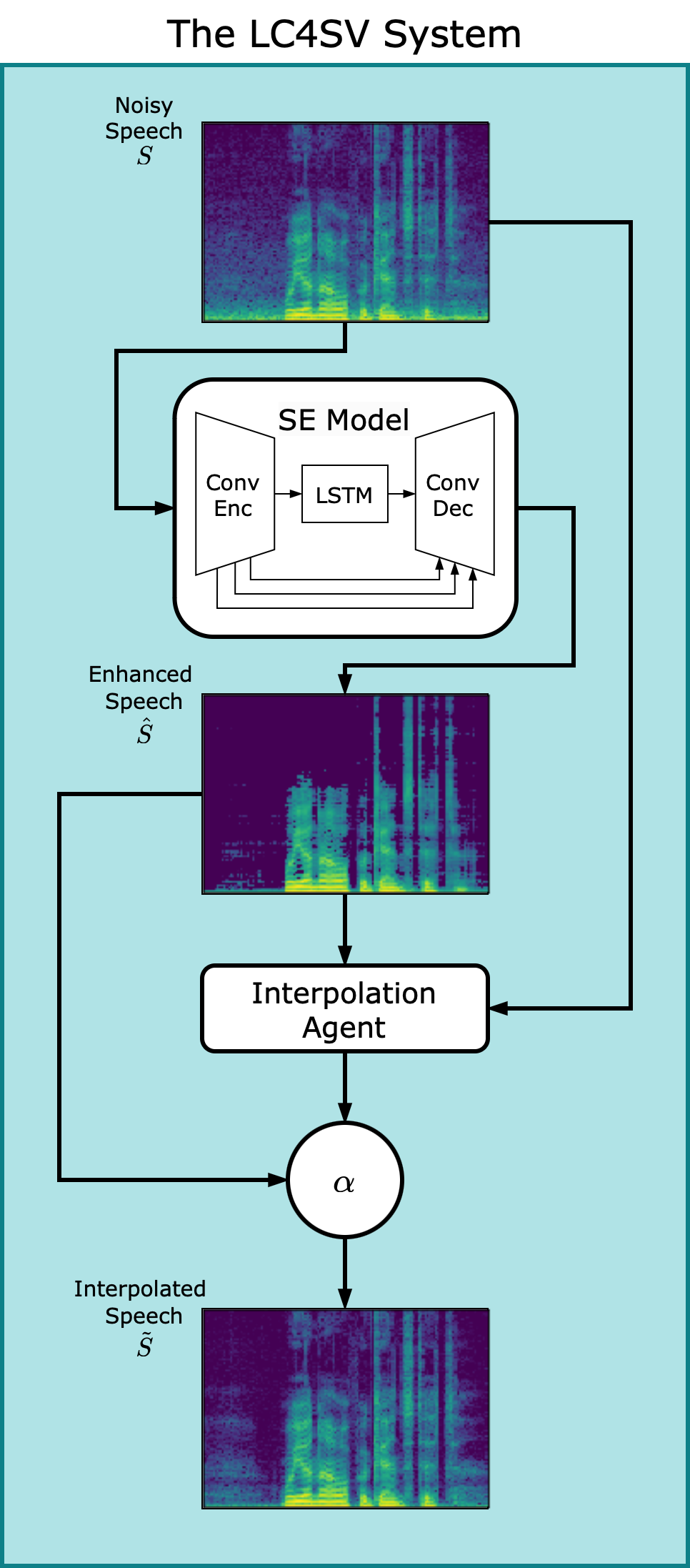}
    \caption{The overall workflow of the LC4SV system.}
    \label{fig:sys}
    \vspace*{-10pt}
\end{figure}

\section{The LC4SV System}
\label{sec:method}
Fig.~\ref{fig:sys} shows the overall workflow of our proposed LC4SV system.
First, referring to VoiceID loss~\cite{Shon2019VoiceIDLS}, we use the SV objective produced by a proxy SV model to train our base SE unit. 
The SE unit converts a noisy speech signal $S$ into an enhanced signal $\hat{S}$.
Next, we train our interpolation agent in an RL manner using an SV-customized reward scheme.
The noisy speech signal and the corresponding enhanced speech signal are concatenated as a two-channel input and fed into the interpolation agent, which determines the appropriate interpolation coefficient $\alpha$.
Finally, we use $\alpha$ to generate a new speech signal $\tilde{S}$ as input to the SV model, where $\tilde{S} = \alpha \hat{S} + (1 - \alpha)S$.
The details of our SE model and interpolation agent are described in Sec.~\ref{sec:se_model} and Sec.~\ref{sec:interpolation_agent}, respectively.

\subsection{Training Stages of the SE Model}
\label{sec:se_model}
We use DEMUCS~\cite{demucs_stft} as our SE model, which consists of an encoder-decoder architecture with skip connections.

\subsubsection{Pre-training}
To provide proper initialization to train the SE model with the SV objective, referring to Alexandre et al.~\cite{demucs_stft}, we pre-train our SE model with a multi-resolution STFT loss to capture information at different time-frequency resolutions.
The objective function $L_{\text{PTN}}$ is defined as:
\begin{equation}
L_{\text{PTN}} = ||\hat{S}-S_{\text{clean}}||_1+ \sum_{i=1}^{M} L_{\text{stft}}^{(i)}(\hat{S}, S_{\text{clean}}),
\end{equation}
where $S_{\text{clean}}$ is the clean target, $M$ is the number of STFT losses with different STFT configurations, $i$ is the configuration index, and $L^{(i)}_{\text{stft}}$ is defined as $L^{(i)}_{\text{sc}}(\hat S,S_{\text{clean}}) + L^{(i)}_{\text{mag}}(\hat S,S_{\text{clean}})$.
$L^{(i)}_{\text{sc}}(\hat S,S_{\text{clean}})$ denotes a spectral convergence loss, and $L^{(i)}_{\text{mag}}(\hat S,S_{\text{clean}})$ denotes a magnitude loss.
$L^{(i)}_{\text{sc}}(\hat S,S_{\text{clean}})$ and $L^{(i)}_{\text{mag}}(\hat S,S_{\text{clean}})$ are respectively defined as:
\begin{equation}
\label{eq:lsc}
L^{(i)}_{\text{sc}}(\hat S,S_{\text{clean}}) = \frac{\||\text{STFT}^{(i)}(\hat{S})| - |\text{STFT}^{(i)}(S_{\text{clean}})|\|_F}{\||\text{STFT}^{(i)}(S_{\text{clean}})|\|_F}
\end{equation}
and
\begin{equation}
\label{eq:lmag}
L^{(i)}_{\text{mag}}(\hat S,S_{\text{clean}}) =\|\log |\text{STFT}^{(i)}(\hat S)| - \log|\text{STFT}^{(i)}(S_{\text{clean}})|\|_1.
\end{equation}

\subsubsection{Fine-tuning}
After pre-training, we fine-tune our SE model using the well-known SV objective - Angular Prototypical (AP)~\cite{chung2020in} loss.
A batch of positive pairs (i.e., two speech signals from the same speaker) is sampled.
Assuming there are $M$ signals of the same speaker in a batch, one signal is randomly selected as the query signal, and the remaining $M - 1$ signals are used to generate the speaker's centroid embedding.
Each speaker's query embedding will be compared to all centroid embeddings to derive similarities for computing the SV objective defined as:
\begin{gather}
\label{eq:ap}
L_{\text{AP}} = -\frac{1}{N}\sum_{i}\log \frac{\exp{(S_{i, c_i})}}{\frac{1}{N}\sum_{j}\exp{(S_{i, c_j})}},
\end{gather}
where $N$ is the batch size (i.e., the number of speakers in the batch), $i$ and $j$ are speaker indices, and $S_{i, c_j}$ is calculated as:
\begin{gather}
\label{eq:ap1}
S_{i, c_j} = w \cdot cos(x_i, c_j) + b,
\end{gather}
where $x_i$ is the $i$-th speaker's query embedding, and 
$c_j$ is the $j$-th speaker's centroid embedding.
$S_{i, c_j}$ is a cosine-based similarity measure with learnable scale $w$ and bias $b$.

In this work, there are 2 enhanced speech signals for each of the 200 speakers in a batch, i.e., $M = 2$ and $N = 200$. 
For each speaker, the embedding of one enhanced speech signal is used as the query, and the embedding of the other enhanced speech signal is used as the centroid.
When updating our SE model, all parameters of the proxy SV model are frozen.
The proxy SV model is a modified version of ResNet-34 (ResNetSE34L) trained with AP loss and provided as the first baseline checkpoint in the GitHub repository of Chung et al.~\cite{chung2020in}, which is publicly available.
\subsection{RL-based Interpolation Agent}
\label{sec:interpolation_agent}
\begin{figure}[h]
        \centering
        \includegraphics[width=0.45\textwidth]{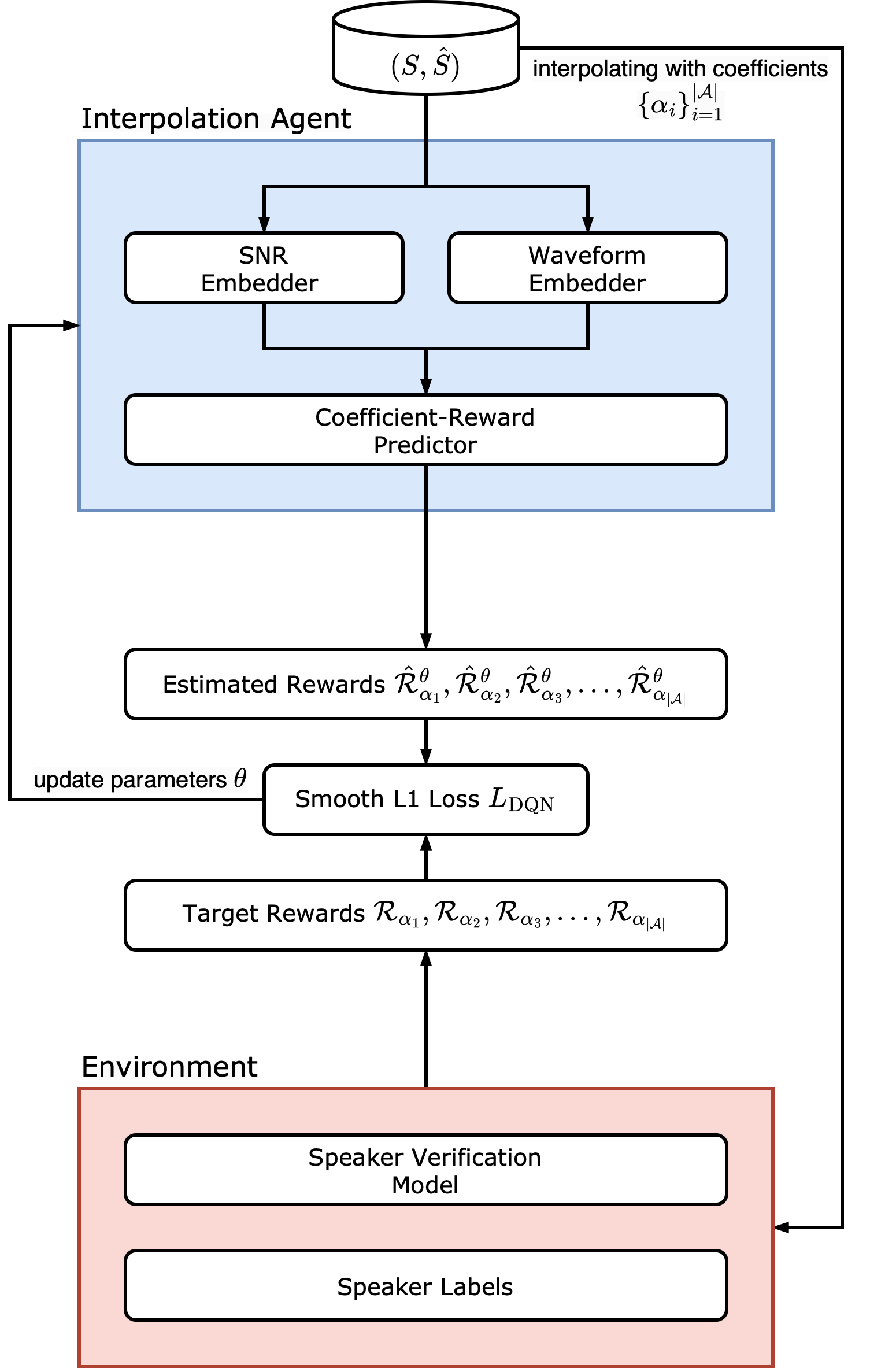}
        \caption{The building pipeline of our interpolation agent.}
        \label{fig:ia}
\vspace*{-10pt}
\end{figure}
The building pipeline of our interpolation agent is shown in Fig.~\ref{fig:ia}.
Our interpolation agent consists of three components: a waveform embedder, an SNR embedder, and a coefficient-reward predictor.
The waveform embedder shares the same feature extraction architecture as the aforementioned proxy SV model, extracting content information from an entire input signal represented as a 256-dimensional embedding.
The SNR embedder maps the SNR value of the noisy signal $S$ to the corresponding SNR representation.
We divide the SNR values into 6 interval groups: $[-\infty, 0)$, $[0, 3)$, $[3, 6)$, $[6, 9)$, $[9, 12)$, and $[12, \infty]$, each corresponding to a specific 256-dimensional SNR embedding.
Note that the 6 SNR embeddings are part of the model parameters of the interpolation agent.
We use the WADA SNR estimation algorithm~\cite{Kim2008RobustSR} to predict the SNR value of the noisy signal $S$.
The coefficient-reward predictor, which employs a multilayer perceptron module with one hidden layer (128 hidden nodes) and the LeakyReLU activation function, predicts the coefficient action based on the waveform embeddings of $S$ and $\hat{S}$ and the SNR embedding.
We denote $\mathcal{A}$ as our action set, $i$ as the coefficient index, and $\alpha_i$ as its element, ranging from 0 to 1 with a step size of 0.1 (11 actions in total).

As in the fine-tuning stage of the SE model, a batch of positive pairs is sampled. Any two speech signals from different speakers within the batch will form a negative pair.
The second baseline model from the GitHub repository of Chung et al.~\cite{chung2020in} is used as the SV model to build an environment that yields rewards from each combination of inputs and actions.
To maximize the effectiveness of the interpolation, we aim to improve discriminability compared to directly inputting the enhanced signal.
Therefore, for the noisy signal $S$ (with speaker embedding $x$), the enhanced signal $\hat{S}$ (with speaker embedding $\hat{x}$), and the coefficient action $\alpha_i$, the reward is defined as:
\begin{equation}
  \begin{aligned}
\mathcal{R}_{\alpha_i}(S, \hat{S}) = \cos(\tilde{x}_{\alpha_i}, \tilde{x}^+_{\alpha_i}) - \cos(\hat{x}, \hat{x}^+) \\
+ \mathbb{E}_{x^-\in \mathcal{N}}[\cos(\hat{x}, \hat{x}^-) - \cos(\tilde{x}_{\alpha_i}, \tilde{x}^-_{\alpha_i})],
\end{aligned}
\end{equation}
where $\tilde{x}_{\alpha_i}$ is the speaker embedding of the interpolated signal $\alpha_i \hat{S} + (1 - \alpha_i)S$;
$\hat{x}^+$ and $\tilde{x}^+_{\alpha_i}$ are the speaker embeddings of the enhanced and interpolated versions of the paired positive sample of $S$;
$\mathcal{N}$ denotes the set of signals of speakers within the batch other than the speaker corresponding to $x$;
${x}^-$, $\hat{x}^-$, and $\tilde{x}^-_{\alpha_i}$ are the speaker embeddings of the noisy, enhanced, and interpolated versions of a negative sample of $S$.
For a positive pair, the similarity between the two interpolated signals (i.e., $\cos(\tilde{x}_{\alpha_i}, \tilde{x}^+_{\alpha_i}$) should be higher than the similarity between the two enhanced signals (i.e., $\cos(\hat{x}, \hat{x}^+)$), i.e., the higher the $\cos(\tilde{x}_{\alpha_i}, \tilde{x}^+_{\alpha_i}) - \cos(\hat{x}, \hat{x}^+)$, the better.
For a negative pair, the similarity between the two enhanced signals (i.e., $\cos(\hat{x}, \hat{x}^-)$) should be higher than the similarity between the two interpolated signals (i.e., $\cos(\tilde{x}_{\alpha_i}, \tilde{x}^-_{\alpha_i})$), i.e., the higher the $\cos(\hat{x}, \hat{x}^-) - \cos(\tilde{x}_{\alpha_i}, \tilde{x}^-_{\alpha_i})$, the better.

To stabilize the training process of the interpolation agent, we choose the Q-Learning algorithm~\cite{dqn} to approximate the action-value function, predicting the improvements made by the selected coefficients and their interpolated signals.
Given the agent's model parameters $\theta$ and the input $(S, \hat{S})$, the output is expressed as $\hat{\mathcal{R}}^{\theta}_{\alpha_i}(S, \hat{S})$.
Here, we employ a smooth L1 loss to compute the distance between the predicted reward and the actual reward. The cost function is expressed as:
\begin{gather}
\label{eq:dqn}
L_{\text{DQN}} = \mathbb{E}_{S, \hat{S}}[\frac{1}{|\mathcal{A}|}\sum_{i=1}^{|\mathcal{A}|} l_{\alpha_i}(S, \hat{S})],
\end{gather}
where $l_{\alpha_i}(S, \hat{S})$ is calculated as:
\begin{gather}
\label{eq:dqn1}
    l_{\alpha_i}(S, \hat{S}) = \begin{cases}
        0.5(e_{\alpha_i}(S, \hat{S}))^2, & \text{if } |e_{\alpha_i}(S, \hat{S})| < 1\\
        |e_{\alpha_i}(S, \hat{S})| - 0.5, & \text{otherwise},
        \end{cases}
\end{gather}
where $e_{\alpha_i}(S, \hat{S})$ is calculated as $\mathcal{R}_{\alpha_i}(S, \hat{S}) - \hat{\mathcal{R}}^{\theta}_{\alpha_i}(S, \hat{S})$.

During inference, we choose $\alpha^*$ as our reacting action according to $\text{argmax}_{\alpha  \in \mathcal{A}} \{\hat{\mathcal{R}}^{\theta}_{\alpha}(S, \hat{S})\}$ to generate the interpolated signal $\tilde{S}$ via $\alpha^* \hat{S} + (1 - \alpha^*)S$.

\begin{table}[h]
    \fontsize{8}{11}\selectfont
    \caption{Performance of different SE architectures evaluated on Libri2Mix.}
    \label{tab:se_model}
    \vspace*{-5pt}
    
    \centering
    \begin{tabular}{l|cccc}
     & STOI & PESQ$_{\text{WB}}$ & PESQ$_{\text{NB}}$ & SI-SDR \\
     \hline
     DCCRN & 0.8857 & 1.8383 & 2.4285 & \textbf{11.9048} \\
     \hline
     DTLN & 0.8577 & 1.6155 & 2.1245 & 10.9175 \\
     \hline
     DEMUCS & \textbf{0.8978} & \textbf{1.9553} & \textbf{2.4486} & 11.2817  
    \end{tabular}
    \vspace*{-15pt}
\end{table}

\begin{table*}[ht]
    \caption{EER (\%) and MinDCF results on Libri2Mix, VOiCES, and the Vox1 test list.}
    \vspace*{-0.5cm}
    \label{tab:exp_SE}
    \setlength{\tabcolsep}{3pt}
    \fontsize{8}{11}\selectfont
    \center
    \subfloat[Ablation I: tested by RawNet3 (RN3) and ECAPA-TDNN (ECP)]{
    \label{tab:baseline}
    \begin{tabular}{|l|cc|cc||cc|cc|} 
\hline
 & \multicolumn{2}{c|}{RN3-Libri2Mix} & \multicolumn{2}{c||}{RN3-VOiCES} & \multicolumn{2}{c|}{ECP-Libri2Mix} & \multicolumn{2}{c|}{ECP-VOiCES} \\ 
\hline
method & EER & MinDCF & EER & MinDCF & EER & MinDCF & EER & MinDCF \\ 
\hline
NOISY & 9.153 & 0.5878 & 2.807 & 0.1317 & 7.964 & 0.4020 & 2.703 & 0.1373  \\
\hline
SE-PTN  & 19.631 & 0.7096 & 13.793 & 0.7739 & 15.890 & 0.6088 & 10.571 & 0.6244 \\
SE-SV  & 9.271 & \textbf{0.4843} & 2.671 & 0.1358 & 7.683 & \textbf{0.3715} & 2.623 & 0.1384 \\
\hline
SE-SV-SNR & 8.877 & 0.4881 & 2.718 & 0.1345 & 7.620 & 0.3818 & 2.633 & 0.1368 \\
LC4SV  & \textbf{8.755} & 0.4873 & \textbf{2.652} & \textbf{0.1345} & \textbf{7.616} & 0.3758 & \textbf{2.621} & \textbf{0.1335} \\
\hline
    \end{tabular}}
    \subfloat[Ablation II: tested by baseline (BSLN)]{
    \label{tab:ef_ia}
    \begin{tabular}{|l|cc|cc|} 
\hline
 & \multicolumn{2}{c|}{BSLN-SV-1} & \multicolumn{2}{c|}{BSLN-SV-2}\\ 
\hline
method & EER & MinDCF & EER & MinDCF\\ 
\hline
NOISY & 2.179 & 0.1691 & 1.018 & 0.0817\\
\hline
SE-PTN & 2.731 & 0.2010 & 1.235 & 0.0883\\
SE-SV  & 2.132 & 0.1671 & 1.023 & 0.0776\\
\hline
SE-SV-SNR & 2.190 & 0.1700 & 1.034 & 0.0824\\
LC4SV  & \textbf{2.073} & \textbf{0.1652} & \textbf{1.018} & \textbf{0.0776}\\
\hline
    \end{tabular}}
\end{table*}

\begin{table*}[ht]
    \vspace*{-5pt}
    \fontsize{8}{11}\selectfont
    \caption{EER (\%) results for different values of $\alpha$ tested using RawNet3 on Libri2Mix.}
    \label{tab:coef}
    \vspace*{-5pt}
    \centering
    \begin{tabular}{l|c|c|c|c|c|c|c|c|c|c}
    $\alpha$ & 0.1 & 0.2 & 0.3 & 0.4 & 0.5 & 0.6 & 0.7 & 0.8 & 0.9 & LC4SV \\
    \hline
     EER (\%) & 9.0235 & 8.9020 & 8.8978 & 8.8140 & 8.8122 & 8.8151 & 8.8811 & 8.9816 & 9.1070 & \textbf{8.7550}\\
    \end{tabular}
    \vspace*{-15pt}
\end{table*}

\section{Experiments}
\label{sec:exp}
\subsection{Experimental Setup and Implementation Details}
\label{exp:implementation}
The training datasets used in the experiments include: 
(1) the DNS-Challenge~\cite{DNS} noise dataset, consisting of 65,303 background and foreground noise samples; 
(2) the Librispeech-960~\cite{panayotov2015librispeech} corpus, consisting of 281,241 utterances; 
(3) the VoxCeleb2~\cite{NAGRANI2020101027} dataset, consisting of 1,092,009 utterances from 5,994 speakers.
The LC4SV framework is divided into three stages: SE model pre-training, SE model fine-tuning, and interpolation agent training.
First, for SE model pre-training, the clean speech utterances from Librispeech-960 were contaminated by randomly sampled noise signals from DNS-Challenge to form noisy-clean pairs for training, with SNR levels uniformly sampled from -4dB to 6dB.
We used the Adam optimizer with $\beta_1$ = 0.9 and $\beta_2$ = 0.999, a learning rate of 0.0002, and a batch size of 8 with 500,000 steps.
Next, with the SV objective derived by the proxy SV model, we fine-tuned the pre-trained SE model using the training data in VoxCeleb2. 
We used the Adam optimizer with $\beta_1$ = 0.9 and $\beta_2$ = 0.999, a learning rate of 0.0001, and a batch size of 32 with 10,000 steps.
Finally, we froze the parameters of the fine-tuned SE model and combined its enhanced signals with its noisy input signals to train our interpolation agent.
The training dataset was also VoxCeleb2.
We used the Adam optimizer with $\beta_1$ = 0.9 and $\beta_2$ = 0.999, a learning rate of 0.0001, and a batch size of 128 with 2,000 steps.

For evaluation, we used two public datasets to verify the effectiveness of our system: LibriMix~\cite{librimix} and noisy VOiCES~\cite{Nandwana2019TheVF, multisv}.
LibriMix took speech utterances from LibriSpeech~\cite{panayotov2015librispeech} and the noise signals from WHAM!~\cite{wham}.
During mixing, the SNR values were sampled from a normal distribution with a mean of 0dB and a standard deviation of 4.1dB.
We tested our LC4SV system on the single-speaker dev and test sets of the Libri2Mix subset, which contains 6,000 utterances.
Noisy VOiCES took speech utterances from the original VOiCES corpus, and the noisy version was produced by MultiSV~\cite{multisv}. 
During mixing, the SNR values were sampled uniformly from 3dB to 20dB.
For each evaluation dataset, we further sampled all speakers to generate pair lists with the same number of positive and negative pairs.
Two unseen SV models were used to evaluate our processing results, RawNet3~\cite{jung2022pushing} and ECAPA-TDNN~\cite{desplanques2020ecapa}.
Both SV models are publicly available and trained on the full VoxCeleb training dataset with additive noise and reverberation augmentation.
More details about them can be found in the GitHub repository of Chung et al.~\cite{chung2020in} and the SpeechBrain toolkit~\cite{speechbrain}.
Additionally, we compared the proposed LC4SV system with several baseline SV systems using the following settings:
(i) directly using the unprocessed noisy speech (termed \textbf{NOISY});
(ii) using the enhanced speech from the SE model pre-trained with the reconstruction loss (termed \textbf{SE-PTN});
(iii) using the enhanced speech from the SE model fine-tuned with the SV objective (termed \textbf{SE-SV});
(iv) selecting the noisy speech or the enhanced speech from the SE model fine-tuned with the SV objective according to a predetermined SNR threshold (termed \textbf{SE-SV-SNR}).
For (iv), the SNR threshold was set to 4dB based on the results on the validation set. 
The SNR values in our experiments were estimated by WADA~\cite{Kim2008RobustSR}.
Two standardized evaluation metrics were used to evaluate the performance: Equal Error Rate (EER [\%]) and Minimum Detection Cost Function (MinDCF). In MinDCF, the prior probability of a target trial $P_{target}$ was set to 0.05. 



\subsection{Comparison of Different SE Models}
\label{sec:se_architecture}
We first validated the choice of DEMUCS as the SE model. 
In addition to DEMUCS, we also trained two widely used SE architectures, DTLN~\cite{dtln} and DCCRN~\cite{hu2020dccrn}, using the same training setup (training data and optimization pipeline) in the pretraining stage. 
The performance in terms of common evaluation scores~\cite{PESQ, STOI, SDR} was evaluated on the LibriMix~\cite{librimix} test set and shown in Table~\ref{tab:se_model}. 
Three standardized evaluation metrics were used to measure SE performance, namely wide/narrow band perceptual evaluation of speech quality ($\text{PESQ}_\text{wb}$/$\text{PESQ}_\text{nb}$)~\cite{PESQ}, short-time objective intelligibility measure (STOI)~\cite{STOI}, and scale-invariant signal-to-distortion ratio (SI-SDR)~\cite{SDR}. 
$\text{PESQ}_\text{nb}$ was designed to evaluate the quality of the processed speech on a scale from -0.5 to 4.5.
STOI was designed to calculate the speech intelligibility on a scale from 0 to 1. 
SI-SDR was designed to measure the energy ratio between speech and non-speech components. 
For all three metrics, higher scores indicate better performance.

The results in Table~\ref{tab:se_model} indicate that while the performance of the three SE models is roughly comparable, DEMUCS exhibits superior performance. These results validate the adequacy of designating DEMUCS as the primary SE architecture for ablation studies of the proposed LC4SV method.

\subsection{Ablation I: Comparison with Baselines}
\label{sec:baseline}
In the first set of ablation experiments, we compared the performance of SV systems operating on different types of speech signals, including noisy (\textbf{NOISY}), enhanced (\textbf{SE-PTN} or \textbf{SE-SV}), either noisy or enhanced according to SNR (\textbf{SE-SV-SNR}), and interpolated (\textbf{LC4SV)} speech.
From Table~\ref{tab:baseline}, several observations can be drawn.
First, \textbf{SE-PTN} dramatically degrades SV performance, indicating that the issue of signal artifacts caused by SE is serious.
Second, \textbf{SE-SV} outperforms \textbf{NOISY} and \textbf{SE-PTN}, indicating the effectiveness of fine-tuning the SE model with the SV objective.
Third, \textbf{SE-SV-SNR} and \textbf{SE-SV} both win and lose in different cases, suggesting that there are advantages and disadvantages to using noisy speech or enhanced speech in SV systems.
Fourth, \textbf{LC4SV} is better than \textbf{SE-SV-SNR} in most cases,
indicating that hard thresholding on the SNR value is not the best strategy for handling distortion compensation.
Fifth, \textbf{LC4SV} is the best method among all compared methods.
Overall, the results validate that our interpolation agent can provide appropriate coefficients to combine enhanced and noisy speech for better SV performance under noisy conditions.


\subsection{Ablation II: Effectiveness of Interpolation Agent}
\label{sec:ablation}
We also evaluated the effectiveness of the interpolation agent in our LC4SV system in an in-domain setting.
Here, the test set was VoxCeleb1~\cite{NAGRANI2020101027}.
The SV systems used for testing were the first and second baseline models from the GitHub repository of Chung et al.~\cite{chung2020in}, which represent the ``seen'' SV models of our LC4SV system, as they were used to fine-tune the SE model and train the interpolation agent, respectively. The results are shown in Table~\ref{tab:ef_ia}.

From Table~\ref{tab:ef_ia}, several observations can be drawn. 
First, all the SE methods except \textbf{LC4SV} degrade the SV performance of BSLN-SV-2. For \textbf{SE-PTN}, \textbf{SE-SV}, and \textbf{SE-SV-SNR}, BSLN-SV-2 is unseen because it only involves interpolation agent training. In contrast, \textbf{SE-SV} brings improvements to BSLN-SV-1 over \textbf{NOISY} since BSLN-SV-1 was seen in \textbf{SE-SV} training. 
Second, we again observe that \textbf{SE-PTN} degrades the performance of BSLN-SV-1 and BSLN-SV-2 due to severe artifacts in its enhanced speech.
Third, \textbf{LC4SV} is the only method that consistently brings improvements over \textbf{NOISY} to both BSLN-SV-1 and BSLN-SV-2  on this in-domain SV task. 
Overall, the results in the in-domain experimental setup again demonstrate the effectiveness of our LC4SV approach, whose interpolation agent effectively determines appropriate coefficients to combine enhanced and noisy speech for better SV performance.

\begin{table}[h]
    \fontsize{8}{11}\selectfont
    \caption{EER (\%) and MinDCF results of RawNet3/ECAPA-TDNN tested on the LibriSpeech dev-clean/test-clean sets.}
    \label{tab:clean}
    \centering
    \begin{tabular}{|l|cc|cc|} 
\hline
 & \multicolumn{2}{c|}{RawNet3} & \multicolumn{2}{c|}{ECAPA-TDNN}\\ 
\hline
method & EER & MinDCF & EER & MinDCF\\ 
\hline
CLEAN & 1.496 & 0.0509 & 1.601 & 0.0614\\
\hline
SE-PTN & 1.408 & 0.0518 & 1.617 & 0.0647\\
SE-SV  & 1.425 & 0.0533 & 1.621 & 0.0633\\
\hline
LC4SV  & 1.383 & 0.0499 & 1.604 & 0.0623\\
\hline
    \end{tabular}
\vspace*{-10pt}
\end{table}
\subsection{Ablation III: Comparison with Predetermined Constant Coefficients}
The key feature of LC4SV is that it automatically and effectively determines the interpolation coefficient $\alpha$ to improve SV performance.
To verify the effectiveness of the sample-by-sample estimation, we compared the performance of determining $\alpha$ using LC4SV with that of using a predetermined value of $\alpha$ ranging from 0.1 to 0.9 with a step size of 0.1. 
The results of testing with the RawNet3 SV model on Libri2Mix are shown in Table~\ref{tab:coef}. From the table, we can notice that \textbf{LC4SV} clearly outperforms other constant coefficients, confirming that \textbf{LC4SV} can effectively estimate the interpolation coefficient.


\subsection{Ablation IV: Oracle Performance Analysis}
In addition to evaluating the performance of our proposed approach in noisy environments, we also conducted experiments using clean speech signals to specifically examine the ability of LC4SV to mitigate artifacts without interference from natural background noise.
To generate a list of verification pairs for testing purposes, we combined the dev-clean and test-clean sets of LibriSpeech. The resulting testing list contained 24,000 positive pairs and 24,000 negative pairs. The results are shown in Table~\ref{tab:clean}.

Comparing the results in Table~\ref{tab:clean} and Table~\ref{tab:baseline}, both SVs perform much better in clean environments than in noisy environments.
The performance drops using the three type of processed signals is marginal compared to clean speech. 
In particular, among the three model settings, LC4SV stands out as the most effective at reducing artifacts and produces results comparable to feeding the clean speech.

\section{Concluding Remarks}
\label{sec:conclusion}
The over-parameterization issue limits speech enhancement models to generate optimal enhanced speech in unknown speaker verification systems.
In this study, we propose a generic denoising framework named LC4SV that learns to compensate for enhanced speech for better generalization.
We use a learning-based interpolation agent to automatically generate appropriate interpolation coefficients between a noisy speech input and its enhanced speech.
Our interpolation agent is trained to maximize the performance of a proxy speaker verification model, preventing information loss in the predictions of speech enhancement models.
Our experimental results demonstrate the effectiveness of LC4SV in mitigating the artifact-induced degradation in speaker verification performance .

\bibliographystyle{IEEEbib}
\bibliography{strings,refs}

\begin{thebibliography}{10}

\bibitem{SE}
Philipos~C. Loizou,
\newblock {\em Speech Enhancement: Theory and Practice},
\newblock CRC Press, Inc., USA, 2nd edition, 2013.

\bibitem{CHLee}
Y.~{Xu}, J.~{Du}, L.~{Dai}, and C.~{Lee},
\newblock ``A regression approach to speech enhancement based on deep neural
  networks,''
\newblock {\em IEEE/ACM Transactions on Audio, Speech, and Language
  Processing}, 2015.

\bibitem{SE1}
M.~{Kolbæk}, Z.~{Tan}, and J.~{Jensen},
\newblock ``Speech intelligibility potential of general and specialized deep
  neural network based speech enhancement systems,''
\newblock {\em IEEE/ACM Transactions on Audio, Speech, and Language
  Processing}, 2017.

\bibitem{SE2}
Zhuo Chen, Shinji Watanabe, Hakan Erdogan, and John~R. Hershey,
\newblock ``Speech enhancement and recognition using multi-task learning of
  long short-term memory recurrent neural networks,''
\newblock in {\em Proc. Interspeech}, 2015.

\bibitem{SE3}
Bingyin Xia and Changchun Bao,
\newblock ``Wiener filtering based speech enhancement with weighted denoising
  auto-encoder and noise classification,''
\newblock {\em Speech Communication}, 2014.

\bibitem{Liu2014ExperimentsOD}
Ding Liu, Paris Smaragdis, and Minje Kim,
\newblock ``Experiments on deep learning for speech denoising,''
\newblock in {\em Proc. Interspeech}, 2014.

\bibitem{DAELu}
Xugang Lu, Yu~Tsao, Shigeki Matsuda, and Chiori Hori,
\newblock ``Speech enhancement based on deep denoising autoencoder,''
\newblock in {\em Proc. Interspeech}, 2013.

\bibitem{7422753}
Xiao-Lei Zhang and DeLiang Wang,
\newblock ``A deep ensemble learning method for monaural speech separation,''
\newblock {\em IEEE/ACM Transactions on Audio, Speech, and Language
  Processing}, pp. 967--977, 2016.

\bibitem{hu2020dccrn}
Yanxin Hu, Yun Liu, Shubo Lv, Mengtao Xing, Shimin Zhang, Yihui Fu, Jian Wu,
  Bihong Zhang, and Lei Xie,
\newblock ``Dccrn: Deep complex convolution recurrent network for phase-aware
  speech enhancement,''
\newblock in {\em Proc. Interspeech}, 2020.

\bibitem{seril}
Chi-Chang Lee, Yu-Chen Lin, Hsuan-Tien Lin, Hsin-Min Wang, and Yu~Tsao,
\newblock ``{SERIL:} noise adaptive speech enhancement using
  regularization-based incremental learning,''
\newblock in {\em Proc. Interspeech}, 2020.

\bibitem{nastar}
Chi-Chang Lee, Cheng-Hung Hu, Yu-Chen Lin, Chu-Song Chen, Hsin-min Wang, and
  Yu~Tsao,
\newblock ``Nastar: Noise adaptive speech enhancement with target-conditional
  resampling,''
\newblock in {\em Proc. Interspeech}, 2022.

\bibitem{pandey2018new}
Ashutosh Pandey and DeLiang Wang,
\newblock ``A new framework for supervised speech enhancement in the time
  domain.,''
\newblock in {\em Proc. Interspeech}, 2018.

\bibitem{1164453}
Y.~Ephraim and D.~Malah,
\newblock ``Speech enhancement using a minimum-mean square error short-time
  spectral amplitude estimator,''
\newblock {\em IEEE/ACM Transactions on Audio, Speech, and Language
  Processing}, pp. 1109--1121, 1984.

\bibitem{SDR}
Jonathan Le~Roux, Scott Wisdom, Hakan Erdogan, and John~R Hershey,
\newblock ``{SDR}-half-baked or well done?,''
\newblock in {\em Proc. ICASSP}, 2019.

\bibitem{demucs_stft}
Alexandre D{\'e}fossez, Gabriel Synnaeve, and Yossi Adi,
\newblock ``Real time speech enhancement in the waveform domain,''
\newblock in {\em Proc. Interspeech}, 2020.

\bibitem{PESQ}
A.~W. {Rix}, J.~G. {Beerends}, M.~P. {Hollier}, and A.~P. {Hekstra},
\newblock ``Perceptual evaluation of speech quality ({PESQ})-a new method for
  speech quality assessment of telephone networks and codecs,''
\newblock in {\em Proc. ICASSP}, 2001.

\bibitem{STOI}
C.~H. {Taal}, R.~C. {Hendriks}, R.~{Heusdens}, and J.~{Jensen},
\newblock ``A short-time objective intelligibility measure for time-frequency
  weighted noisy speech,''
\newblock in {\em Proc. ICASSP}, 2010.

\bibitem{reddy2021dnsmos}
Chandan~KA Reddy, Vishak Gopal, and Ross Cutler,
\newblock ``Dnsmos: A non-intrusive perceptual objective speech quality metric
  to evaluate noise suppressors,''
\newblock in {\em Proc. ICASSP}, 2021.

\bibitem{cooper2022generalization}
Erica Cooper, Wen-Chin Huang, Tomoki Toda, and Junichi Yamagishi,
\newblock ``Generalization ability of mos prediction networks,''
\newblock in {\em Proc. ICASSP}, 2022.

\bibitem{d4am}
Chi-Chang Lee, Yu~Tsao, Hsin-Min Wang, and Chu-Song Chen,
\newblock ``D4{AM}: A general denoising framework for downstream acoustic
  models,''
\newblock in {\em Proc. ICLR}, 2023.

\bibitem{espnet}
Yen-Ju Lu, Xuankai Chang, Chenda Li, Wangyou Zhang, Samuele Cornell, Zhaoheng
  Ni, Yoshiki Masuyama, Brian Yan, Robin Scheibler, Zhong-Qiu Wang, et~al.,
\newblock ``Espnet-se++: Speech enhancement for robust speech recognition,
  translation, and understanding,''
\newblock in {\em Proc. Interspeech}, 2022.

\bibitem{9747146}
Yen-Ju Lu, Samuele Cornell, Xuankai Chang, Wangyou Zhang, Chenda Li, Zhaoheng
  Ni, Zhong-Qiu Wang, and Shinji Watanabe,
\newblock ``Towards low-distortion multi-channel speech enhancement: The
  espnet-se submission to the l3das22 challenge,''
\newblock in {\em Proc. ICASSP}, 2022.

\bibitem{deto}
Seyed Sadjadi and John Hansen,
\newblock ``Assessment of single-channel speech enhancement techniques for
  speaker identification under mismatched conditions,''
\newblock in {\em Proc. Interspeech}, 2010.

\bibitem{9064910}
Hassan Taherian, Zhong-Qiu Wang, Jorge Chang, and DeLiang Wang,
\newblock ``Robust speaker recognition based on single-channel and
  multi-channel speech enhancement,''
\newblock {\em IEEE/ACM Transactions on Audio, Speech, and Language
  Processing}, pp. 1293--1302, 2020.

\bibitem{Shon2019VoiceIDLS}
Suwon Shon, Hao Tang, and James~R. Glass,
\newblock ``Voiceid loss: Speech enhancement for speaker verification,''
\newblock in {\em Proc. Interspeech}, 2019.

\bibitem{9747771}
Ladislav Mošner, Oldřich Plchot, Lukáš Burget, and Jan~Honza Černocký,
\newblock ``Multi-channel speaker verification with conv-tasnet based
  beamformer,''
\newblock in {\em Proc. ICASSP}, 2022.

\bibitem{10022350}
Sandipana Dowerah, Romain Serizel, Denis Jouvet, Mohammad Mohammadamini, and
  Driss Matrouf,
\newblock ``Joint optimization of diffusion probabilistic-based multichannel
  speech enhancement with far-field speaker verification,''
\newblock in {\em Proc. SLT}, 2023.

\bibitem{chung2020in}
Joon~Son Chung, Jaesung Huh, Seongkyu Mun, Minjae Lee, Hee~Soo Heo, Soyeon
  Choe, Chiheon Ham, Sunghwan Jung, Bong-Jin Lee, and Icksang Han,
\newblock ``In defence of metric learning for speaker recognition,''
\newblock in {\em Proc. Interspeech}, 2020.

\bibitem{NAGRANI2020101027}
Arsha Nagrani, Joon~Son Chung, Weidi Xie, and Andrew Zisserman,
\newblock ``Voxceleb: Large-scale speaker verification in the wild,''
\newblock {\em Computer Speech \& Language}, p. 101027, 2020.

\bibitem{zeng2022joint}
Chang Zeng, Xiaoxiao Miao, Xin Wang, Erica Cooper, and Junichi Yamagishi,
\newblock ``Joint speaker encoder and neural back-end model for fully
  end-to-end automatic speaker verification with multiple enrollment
  utterances,''
\newblock {\em arXiv preprint arXiv:2209.00485}, 2022.

\bibitem{enh_not}
Hiroshi Sato, Tsubasa Ochiai, Marc Delcroix, Keisuke Kinoshita, Naoyuki Kamo,
  and Takafumi Moriya,
\newblock ``Learning to enhance or not: Neural network-based switching of
  enhanced and observed signals for overlapping speech recognition,''
\newblock in {\em Proc. ICASSP}, 2022.

\bibitem{Kim2008RobustSR}
Chanwoo Kim and Richard~M. Stern,
\newblock ``Robust signal-to-noise ratio estimation based on waveform amplitude
  distribution analysis,''
\newblock in {\em Proc. Interspeech}, 2008.

\bibitem{dqn}
Volodymyr Mnih, Koray Kavukcuoglu, David Silver, Andrei Rusu, Joel Veness, Marc
  Bellemare, Alex Graves, Martin Riedmiller, Andreas Fidjeland, Georg
  Ostrovski, Stig Petersen, Charles Beattie, Amir Sadik, Ioannis Antonoglou,
  Helen King, Dharshan Kumaran, Daan Wierstra, Shane Legg, and Demis Hassabis,
\newblock ``Human-level control through deep reinforcement learning,''
\newblock {\em Nature}, pp. 529--33, 2015.

\bibitem{DNS}
Chandan~KA Reddy, Vishak Gopal, Ross Cutler, Ebrahim Beyrami, Roger Cheng,
  Harishchandra Dubey, Sergiy Matusevych, Robert Aichner, Ashkan Aazami,
  Sebastian Braun, et~al.,
\newblock ``The interspeech 2020 deep noise suppression challenge: Datasets,
  subjective testing framework, and challenge results,''
\newblock in {\em Proc. Interspeech}, 2020.

\bibitem{panayotov2015librispeech}
Vassil Panayotov, Guoguo Chen, Daniel Povey, and Sanjeev Khudanpur,
\newblock ``Librispeech: an asr corpus based on public domain audio books,''
\newblock in {\em Proc. ICASSP}, 2015.

\bibitem{librimix}
Joris Cosentino, Manuel Pariente, Samuele Cornell, Antoine Deleforge, and
  Emmanuel Vincent,
\newblock ``Librimix: An open-source dataset for generalizable speech
  separation,''
\newblock in {\em Proc. Interspeech}, 2020.

\bibitem{Nandwana2019TheVF}
Mahesh~Kumar Nandwana, Julien van Hout, Colleen Richey, Mitchell McLaren,
  Maria~Alejandra Barrios, and Aaron~D. Lawson,
\newblock ``The voices from a distance challenge 2019,''
\newblock in {\em Proc. Interspeech}, 2019.

\bibitem{multisv}
Ladislav Mošner, Oldřich Plchot, Lukáš Burget, and Jan Černocký,
\newblock ``Multisv: Dataset for far-field multi-channel speaker
  verification,''
\newblock in {\em Proc. ICASSP}, 2022.

\bibitem{wham}
Gordon Wichern, Joe Antognini, Michael Flynn, Licheng~Richard Zhu, Emmett
  McQuinn, Dwight Crow, Ethan Manilow, and Jonathan~Le Roux,
\newblock ``Wham!: Extending speech separation to noisy environments,''
\newblock {\em arXiv preprint arXiv:1907.01160}, 2019.

\bibitem{jung2022pushing}
Jee-weon Jung, You~Jin Kim, Hee-Soo Heo, Bong-Jin Lee, Youngki Kwon, and
  Joon~Son Chung,
\newblock ``Pushing the limits of raw waveform speaker recognition,''
\newblock in {\em Proc. Interspeech}, 2022.

\bibitem{desplanques2020ecapa}
Brecht Desplanques, Jenthe Thienpondt, and Kris Demuynck,
\newblock ``{ECAPA-TDNN: Emphasized Channel Attention, propagation and
  aggregation in TDNN based speaker verification},''
\newblock in {\em Proc. Interspeech}, 2020.

\bibitem{speechbrain}
Mirco Ravanelli, Titouan Parcollet, Peter Plantinga, Aku Rouhe, Samuele
  Cornell, Loren Lugosch, Cem Subakan, Nauman Dawalatabad, Abdelwahab Heba,
  Jianyuan Zhong, Ju-Chieh Chou, Sung-Lin Yeh, Szu-Wei Fu, Chien-Feng Liao,
  Elena Rastorgueva, François Grondin, William Aris, Hwidong Na, Yan Gao,
  Renato De~Mori, and Yoshua Bengio,
\newblock ``Speechbrain: A general-purpose speech toolkit,''
\newblock {\em arXiv preprint arXiv:2106.04624}, 2021.

\bibitem{dtln}
Nils Westhausen and Bernd Meyer,
\newblock ``Dual-signal transformation lstm network for real-time noise
  suppression,''
\newblock in {\em Proc. Interspeech}, 2020.

\end{thebibliography}

\end{document}